\documentclass[aps,prl,twocolumn]{revtex4}
\usepackage{amsfonts}
\usepackage{amsmath}
\usepackage{amssymb}
\usepackage{epsfig}
\usepackage{graphicx}
\begin{document}

\title{Dipolar gases in quasi one-dimensional geometries}
\author{S. Sinha$^{1}$ and L. Santos$^2$} 
\affiliation{ 
\mbox{$^1$ S. N. Bose National Centre for Basic Sciences, 
700098 Kolkata, India} \\
\mbox{$^2$ Institut f\"ur
Theoretische Physik, Leibniz Universit\"at Hannover,
Appelstr. 2, D-30167 Hannover, Germany}\\
}

\begin{abstract}
We analyze the physics of cold dipolar gases in quasi one-dimensional 
geometries, showing that the confinement-induced scattering resonances 
produced by 
the transversal trapping are crucially affected by the dipole-dipole 
interaction. As a consequence, the dipolar interaction may drastically change 
the properties of quasi-1D dipolar condensates, even for situations 
in which the dipolar interaction would be completely overwhelmed 
by the short-range interactions in a 3D environment. 
\end{abstract}
\maketitle




Low-dimensional ultra cold gases have recently attracted a major attention. 
One and two-dimensional gases are created in sufficiently strong  
optical lattices \cite{Esslinger-Old}, or by means of magnetic wires  
\cite{Schmiedmayer}. Low-dimensionality leads to a very rich physics, 
highlighted by the recently observed Berezinskii-Kosterlitz-Thouless 
transition in 2D gases \cite{Dalibard-BKT}, the enhanced role of thermal 
and quantum phase fluctuations in elongated gases 
\cite{Phase-fluctuations}, and the realization of the 
Tonks-Girardeau regime of 1D bosons 
\cite{Tonks}. 


The properties of quantum gases are crucially determined by the interparticle
interactions. Current experiments typically involve particles at very low
energies interacting via a short-range 
isotropic potential characterized by an $s$-wave scattering length $a$. 
The latter may be modified by means of scattering resonances 
induced by magnetic fields (Feshbach resonance) or by 
properly detuned lasers \cite{Feshbach}.  
Interestingly, the scattering properties in quasi-1D  
(also in 2D \cite{Petrov}) may be crucially affected by the contrained 
geometry. 
In particular the transversal confinement can induce a novel 
resonance known as confinement-induced resonance (CIR) \cite{Olshanii}, 
in which the transversal trap modes assume similar roles as the 
open and closed channels in usual Feshbach resonances \cite{Olshanii2}.


Interestingly, new experiments on atoms with large magnetic moment \cite{Chromium}
cold molecules \cite{Molecules} and Rydberg atoms \cite{Rydberg} 
have recently opened a fascinating new research area, namely that of dipolar
gases. In these gases, the dipole-dipole interaction (DDI), which is
long-range and anisotropic, plays a significant or even dominant role when 
compared to the short-range isotropic interaction. The special features of the
DDI lead to fundamentally new physics in Bose-Einstein condensates 
\cite{YiYou,Dip-BEC}, degenerated Fermi gases \cite{Dip-Ferm}, 
strongly-correlated atomic systems \cite{Dip-Str}, 
quantum computation \cite{Dip-QInf}, and ultra cold chemistry \cite{Dip-Chem}.
Recently, time-of-flight experiments with Chromium condensates have allowed for the first observation ever  
of dipolar effects in quantum gases  \cite{Expansion}.


To a very good approximation the combined effects of the short-range interactions and the DDI 
can be understood by means of a pseudopotential theory, which includes a contact interaction, 
characterized by a scattering length $a$, and the DDI itself 
\cite{YiYou,Ronen2006}. However, the correct value of $a$ is 
in general not that in absence of DDI, but  
the result of the scattering problem including both the
short-range and the DDI potentials. Hence the DDI may affect the
value of $a$, even very severely in the vicinity of the so-called shape
resonances  \cite{YiYou,Ronen2006,Wang2006}. In addition, the combination of the DDI
and the dressing of rotational excitations with static 
and microwave fields may allow for the engineering of 
novel types of interaction potentials for polar molecules in 2D geometries \cite{Buchler2006}.


In this Letter, we analyze the scattering properties of dipolar gases in quasi-1D geometries. 
By solving the corresponding scattering problem, including the short-range
interaction, the DDI, and the trap potential, we obtain an effective
1D pseudopotential consisting of a contact interaction with an
effective 1D coupling constant and a regularized dipolar
potential. Similar as for the case without DDI we observe the appearance of a
CIR, however the position of the CIR is crucially modified by the DDI. In
particular, even for large values of $a$ for which in a 3D environment  
the short-range interaction fully dominates the DDI, the latter can  
dramatically modify the 1D scattering. As a consequence, the properties of the
1D gas may be crucially modified, leading to observable effects, which we
discuss at the end of this Letter.



In the following, we consider a dipolar gas of particles with (electric or
magnetic) dipole moment $d$ in a quasi-1D geometry along the axial $x$-direction. 
The transversal $yz$-confinement is given by an harmonic potential of frequency $\omega$, whereas no 
trapping is assumed in the $x$-direction.  We are interested in the scattering of two 
dipolar particles under these conditions, and hence (due to the separability of the Hamiltonian) 
we consider the corresponding scattering problem
in the relative coordinate $\vec r$, given by the Schr\"odinger equation:
\begin{equation}
[-\frac{\hbar^2}{2\mu} \nabla^2 + \frac{1}{2}\mu \omega^2 \rho^2 +V_{sh}(\vec{r}) + V_{d}(\vec{r})]\psi = E \psi,
\end{equation}
where $\rho^2=y^2+z^2$, $\mu = m/2$ is the relative mass, $V_{sh}(\vec{r})$ is the short-range potential, and 
$E = \hbar \omega + \hbar^2 k^2/2\mu$, with $k$ the axial momentum. 
Assuming the dipoles oriented along $x$, the DDI is given by:
\begin{equation}
V_d (\vec{r})= \frac{d^2}{r^3}\left[1 - \frac{3 x^2}{r^2}\right]
\label{schreq}
\end{equation}
and hence the DDI is attractive along $x$. This configuration 
is particularly convenient for the analysis 
since it maintains the cylindrical symmetry of the problem. 
This symmetry may be also preserved if the dipoles while forming an angle $\phi$ with the $x$-axis 
are rotated in the plane perpendicular to the trap axis fast enough to lead to an effective 
time averaged DDI \cite{Tuning}, similar to that of Eq.~(\ref{schreq}) but in
which $d^2$ transforms into $\alpha d^2$, where $\alpha$ may range between $1$ ($\phi=0$) and 
$-1/2$ ($\phi=\pi/2$). Note that for $\alpha<0$ the DDI becomes repulsive along the axis. 

If the system is assumed in the single-mode approximation (SMA), i.e. 
only the transversal ground-state is considered, 
the wavefunction may be split as 
$\psi(x,\rho)=f(x) \phi_{0}(\rho)$, where 
$\phi_{0}(\rho) = \exp [-\rho^2/2 l^2]/\sqrt{\pi} l$, 
with  $l^2 = \hbar/(\mu \omega)$. After integrating 
over the transversal coordinates, the DDI leads to a delta-like 
interaction plus an effective 1D DDI of the form:
\begin{equation}
V_{1d} (x)= \frac{2\alpha d^2}{l^3} \left[2 \sqrt{t}- 
\sqrt{\pi}(1 + 2 t)e^{t} {\rm erfc}[\sqrt{t}] \right]
\label{V1d}
\end{equation}
where $t = (x/l)^2$, {\rm erfc} denotes the complementary error function. 
For distances $x\gg l$, $V_{1d}\propto 1/x^3$ as expected. 
Note that although the DDI is long-ranged and divergent at the origin, $V_{1d}(x)$ 
is regularized at $x=0$ as well as in momentum space, being 
finite for $k =0$, leading us to use (\ref{V1d}) in the pseudopotential below.


To formulate an effective theory for the low-energy scattering of the dipolar particles 
a proper pseudopotential must be introduced, which as shown below  
may be chosen as a combination of a regularized 1D DDI and an appropriate 
contact potential. The motivation for this choice may be understood 
by considering, still within the SMA, the scattering of two dipoles 
at a short-range potential of range $b$.   
For $d=0$, the psedupotential is given by $g_{1D} \delta(x)$, where $g_{1D}$
is a given coupling constant. However, for $d\neq 0$, 
the DDI dominates the asymptotic behavior of the wavefunction for $x >> b$, 
which for $k=0$ is given by the equation
$d^2\psi/dx^2 = -(2\alpha \mu  d^2/\hbar^2) \psi/|x|^3$, whose 
general solution (for $\alpha=1$) is: 
\begin{equation}
\psi(x) = \sqrt{x}\left[A J_{1}(2\sqrt{2l_d/x}) +
B Y_{1}(2\sqrt{2l_d/x})\right]
\end{equation}
where $A$ and $B$ are constants, $J_1$ and $Y_1$ are Bessel functions, and 
$l_d = d^2 \mu/\hbar^2$ is a length scale associated to the DDI. Asymptotically,  
$\psi(x\rightarrow \infty) = A'\left[x/l_d + 2\log(x/l_d)\right] + B'$, 
where $A'$ and $B'$ are constants. Note that the asymptotic form 
of the wavefunction is not linear, and therefore   
the true potential cannot be replaced solely by a contact pseudopotential, as
for $d=0$ \cite{footnote0}. We choose thus a convenient pseudopotential providing 
the correct asymptotic behavior, formed by a contact interaction, and 
the regularized DDI (\ref{V1d}):
\begin{equation}
v_{eff}  =  g_{1D}\delta(x) + V_{1d}(x)
\end{equation}
where $g_{1D}$ is obtained for a particular short-range potential,
after matching the logarithmic derivative at $x = b$.


The SMA may become insufficient to describe the 
two-body scattering problem in a quasi-1D geometry. In particular, as discussed above, 
an analysis beyond the SMA \cite{Olshanii,Olshanii2} shows that for short-range
interactions the scattering process in the presence of a transverse harmonic confinement 
potential undergoes a CIR, at which the coupling strength of the effective 1D pseudopotential diverges. 
For a 3D contact pseudopotential 
$\frac{4\pi \hbar^2 a}{m} \delta(\vec{r})$, the effective 1D coupling constant is given by \cite{Olshanii}:
\begin{equation}
g_{1D} = \frac{2 \hbar^2 a}{\mu l^2}\frac{1}{(1 - 1.46 a/l)},
\label{CIR}
\end{equation}
and hence $g_{1D}$ diverges when $a \sim 0.68 l$.
It has been verified numerically that the shape and the position
of the CIR does not depend on the details of the interatomic interaction, 
when the range of the interaction is much smaller than the radial confinement length \cite{Olshanii2}.


In the following we analyze the effects of the dipolar interaction in the CIR  
by considering the scattering process beyond the SMA. We solve the 3D scattering 
(simplified by the cylindrical symmetry of the problem) numerically. 
For simplicity, we consider a simplified model 
for the short range potential, namely  
a finite-depth potential well: $V_{sh} = -V_{0} \theta(r_{0} - r)$, 
for which the 3D scattering length $a$ is
analytically known $a = r_{0}\left[1 - \tan(\sqrt{mV_{0}}r_{0}/\hbar)/\sqrt{mV_{0}}
r_{0}/\hbar\right]$ \cite{footnote-PT}. 
$r_0$ denotes the range of the interactions, which is kept small compared to the 
radial confinement length $l$ (in the calculations below we use $r_{0} = 0.1 l$). 
In the following we assume that for the DDI considered the system is
sufficiently far away from shape resonances, and hence we maintain the
analytical value of $a$ obtained in absence of the DDI. 
The 3D dipolar interaction is considered of the form
$\frac{\alpha d^2}{r^2}\left[1 - \frac{3 x^2}{r_{\perp}^2 + x^2}\right]\theta(r - r_{0})$, 
where we have imposed a cut-off 
at short distances to avoid divergences. This cut-off is physically justified
since at short distances the short-range potential dominates.

To evaluate the effective 1D coupling constant $g_{1D}$ we solve the 3D 
Schr\"odinger equation~(\ref{schreq}) for a momentum $k=0$. 
Employing the cylindrical symmetry of the problem, we consider a 
2D numerical grid for the radial and axial coordinates, 
choosing the maximal radial value such that the wave function vanishes 
at the border, whereas the axial box has a length  
$L>> l>>l_{d} >>r_{0}$. Note that the latter restricts our calculation to 
sufficiently small values of dipole moments to satisfy the 1D condition $l>l_{d}$.
In order to increase the precision near the scattering center, we employ 
a non-uniform grid $\eta(i)=$sech$(\Delta_\eta i)$, where $\eta=\rho,x$, and 
$\Delta_\eta$ are properly chosen.
Since only the even parity wave function contributes to the 
low-energy scattering problem, we impose the boundary condition
$\frac{\partial \psi(0, \rho)}{\partial x} = 0$.
For $L\gg l_d$, the wave function can
be written in a product form: $\psi(x,\rho) = f(x)\phi_{0}(\rho)$.
Imposing at $x=L$ the logarithmic derivative $\frac{1}{f}\frac{df}{dx}$ in a fully 1D calculation, 
we evolve the 1D wavefunction from $x=L$ to $x=0$ using $V_{1d}(x)$. 
From the logarithmic derivative of the 1D wavefunction at $x=0$ we obtain $g_{1D}$.

\begin{figure}
\begin{center}
\includegraphics[width=6.0cm]{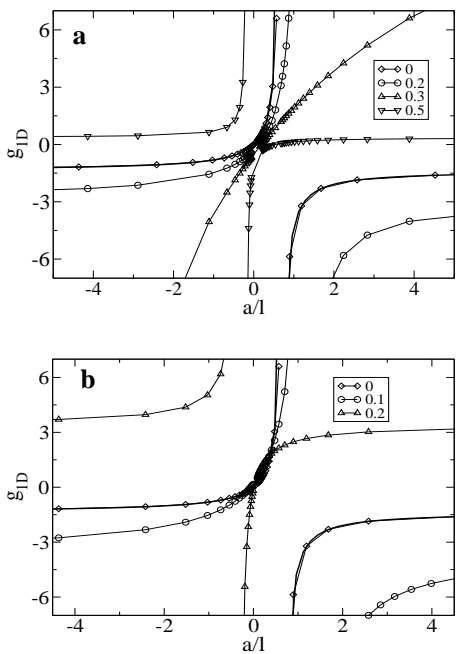}
\vspace*{-0.3cm} 
\end{center} 
\caption{
$g_{1D}$ in units of $\hbar^2/\mu l$ as a function of $a/l$ for various values of $d^2/l^3\hbar\omega$ 
for (a) $\alpha=-1/2$ (repulsive dipoles), and (b) $\alpha=1$ (attractive dipoles). 
The bold line represents the analytical results of Ref.~\cite{Olshanii}.}
\label{fig:1}  
\end{figure}

Fig.~\ref{fig:1} shows the value of $g_{1D}$ 
for different dipolar strengths $d^2/l^3\hbar\omega$, as 
a function of the 3D scattering length.  
For vanishing dipole moment, our numerical results are in
good agreement with the analytical formula~(\ref{CIR})~\cite{Olshanii}.
For sufficiently large values of $a$, the effective coupling constant 
approaches an asymptotic universal negative value
$-\frac{1.37\hbar^2}{\mu l}$.
The DDI significantly modifies the behavior of $g_{1D}$ as a function of $a$. 
With growing $d$ the position of the CIR shifts towards 
larger positive values of $a$. At some particular value of the dipole
strength ($d^2/l^3 \sim 0.3\hbar \omega$), the position 
of the CIR is displaced towards $a=+\infty$, and as 
a consequence the CIR dissappears and $g_{1D}$ 
monotonically increases with $a$. Further increasing $|d|$ 
shifts the CIR to negative values of $a$. When increasing $d$ even further the CIR 
is found again for $a>0$, scanning again all values of $a$
until $+\infty$ and back from $-\infty$ to $0$, and so on in a cyclic way.


Note that in SMA a sufficiently large repulsive DDI on the $x$-axis would lead eventually to a shielding of the 
short-range potential, and $g_{1D}$ would become an universal function of $d$. However, in the calculations 
presented in this paper we consider the case in which $l\gg l_d$. Therefore the anisotropy of the DDI becomes important, 
and the relative wavefunction can surround the repulsive axial barrier, and explore the short-range part of the potential.

\begin{figure}
\begin{center}
\includegraphics[width=5.0cm]{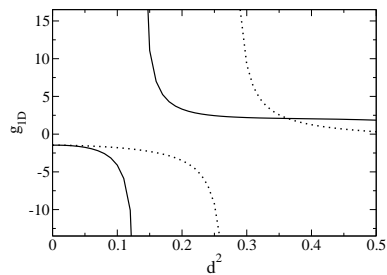}
\vspace*{-0.3cm}
\end{center}
\caption{ 
$g_{1D}$ in units of $\hbar^2/\mu l$ as a function of $d^2$ in units of $l^3\hbar\omega$ 
for a fixed 3D scattering length $a=10.1 l$, for 
$\alpha=-1/2$ (dotted line) and $\alpha=1$ (solid line).}
\label{fig:2}
\end{figure}
\begin{figure}[ht]
\begin{center}
\includegraphics[width=5.0cm]{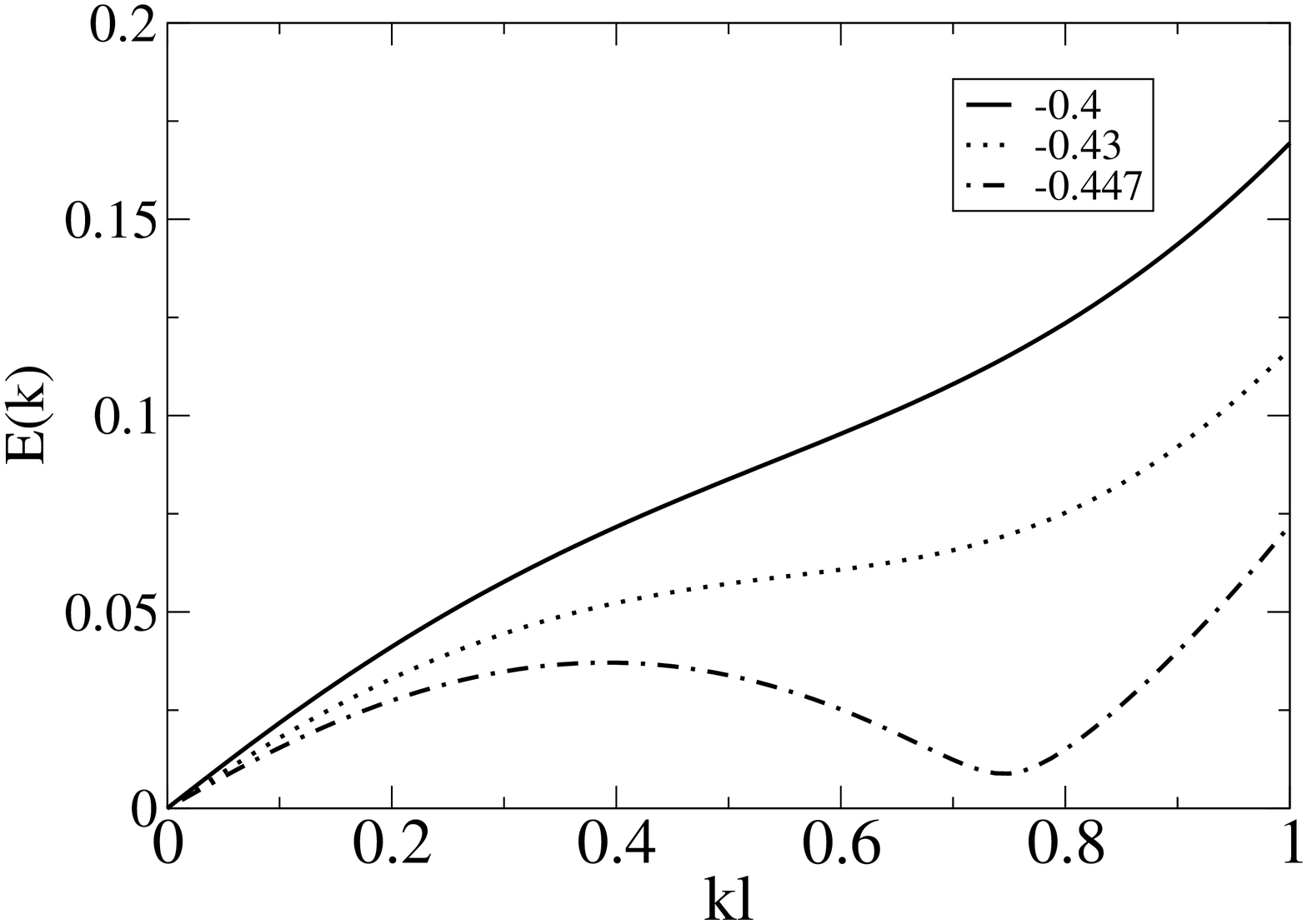}
\vspace*{-0.4cm}
\end{center}
\caption{ Quasi particle excitation energy $E(k)$ in units of $\hbar \omega$ as a function of $k l$, for 
 $2nd^{2}/l^{2}= 0.5 \hbar \omega$, for  
$n g_{1D}/\hbar \omega = -0.4$ (solid), $-0.43$ (dashed), and $-0.447$ (dotted-dashed).}
\label{fig:3}
\vspace*{-0.1cm}
\end{figure}

Note that in a 3D environment when $|a|$ is sufficiently large the DDI becomes 
irrelevant compared to the short-range interactions. Remarkably, this
apparently intuitively obvious fact is not what occurs in a waveguide geometry.
Fig.~\ref{fig:2} shows the asymptotic value of $g_{1D}$ for large $|a|$, which 
interestingly, due to the DDI can approach a positive value,
i.e. the gas acquires a repulsive character. This qualitatively differs from
the behavior for $d=0$, for which the effective 1D contact
interactions acquire for sufficiently large $|a|$ an universal attractive
character \cite{Olshanii}. Hence, even if the short-range interactions fully dominate 
the physics in a 3D geometry, the DDI can dramatically change the properties of the dipolar gas 
in quasi-1D geometries, eventually changing the sign of $g_{1D}$.
The prediction of this remarkable property introduced by the DDI 
(which should be experimentally observable for 
quasi-1D dipolar gases at a Feshbach resonance)  
can be considered the main result of this Letter. 

Once the pseudopotential is known,  
one writes down the many-body Hamiltonian for quasi-1D dipolar bosons
\begin{eqnarray}
& &  H   =  \int dx \Psi^{\dagger}(x)\left[\frac{-\hbar^2}{2m}
\frac{\partial^2}{\partial x^2} +
 \frac{1}{2}g_{1D}\Psi^{\dagger}(x)\Psi (x)\right]\Psi(x) \nonumber\\
& + & \iint dx dx' \Psi^{\dagger}(x)\Psi^{\dagger}(x') 
\frac{V_{1d}(|x-x'|)}{2} \Psi(x)\Psi(x')
\end{eqnarray}
where $\Psi(x)$ ($\Psi^{\dagger}(x)$) is the creation (anihilation) operator of
bosons at $x$. At the CIR a Tonks gas with additional DDI would lead to a 
super-Tonks gas, with Luttinger parameter $K<1$ \cite{giamarchi}.  
For small values of $|g_{1D}|$, and although strictly condensation is
prevented in 1D, a finite-size system at a sufficiently low temperature 
allows for a quasi-condensate  \cite{phase-fluctuation}. This quasi-1D 
dipolar BEC 
may present a remarkable physics, as it becomes clear from an analysis of the 
dispersion law $E(k)$ for axial excitations of momentum $k$ on top of an 
homogeneous quasi-1D BEC:
\begin{equation}
E(k) = \sqrt{\epsilon(k)\left[\epsilon(k) + 
2 \left (g_{1D} - 
\tilde{V}_{1d}(k) \right ) n\right]} 
\end{equation}
where $\epsilon(k)=\hbar^2k^2/2m$, $\tilde{V}_{1d}(k) =  \frac{4\alpha d^2}{l^2}\left[ 1 - 
\sigma e^{\sigma} \Gamma(0,\sigma)\right]$ is the 
Fourier transform of $V_{1d}(x)$, with $\sigma=k^2l^2/4$.  
Note that $|\tilde{V}_{1d}(k)|$ monotonically decreases with 
$k$, with $|\tilde{V}_{1d}(0)|=4|\alpha| d^2/l^2$. Hence, phonon stability 
demands $g_{1D}>V_{1D}(0)$, since otherwise the homogeneous 
quasi-1D BEC becomes unstable against the formation of bright solitons. 
However, even for $g_{1D}>V_{1D}(0)$, 
the homogeneous BEC can become eventually unstable. This occurs 
when $g_{1D}<0$, and $\alpha<0$ (repulsive DDI along the axis). 
In that case the function $g_{1D}-\tilde V_{1d}(k)$ changes 
its sign for a sufficiently large $k$, leading to the possibility 
of achieving a roton (see Fig.~\ref{fig:3}) 
in the spectrum at intermediate values of $k$ 
\cite{gora,Roton1D,sinha}. When $|g_{1D}|$ increases, the roton becomes deeper, 
and for a critical $g_{1D}=g_c$, the roton minimum  
reaches zero energy at a finite momentum $k_{c}$, leading to an instability 
of the homogeneous quasi-1D BEC.  
For $\alpha=-1/2$, and a given $d$, $k_c$ may be obtained from the equation 
$(1 + \sigma_c)e^{\sigma_c}\Gamma[0,\sigma_c] = 
1 + l^2 \hbar \omega/(4 d^2 n)$, where $\sigma_c = k_{c}^2 l^2/4$, whereas 
$-2g_{c} n/\hbar \omega = (1 - \sigma_c + \sigma_c^{2}e^{\sigma_c}
\Gamma[0,\sigma_c])/(1 - (1 + \sigma_c)e^{\sigma_c}\Gamma[0,\sigma_c])$.

In summary, the DDI may play a crucial role in the physics of 
quasi-1D condensates, even for situations for which the short 
range interactions overwhelm the DDI in a 3D environment. 
The properties of the quasi-1D dipolar BEC crucially 
depend on the value of $g_{1D}$, which in turn will depend in 
a non-trivial way on the dipole and dipole-orientation due to the 
dipole-induced modification of the CIR. In particular, the 
DDI may change the sign of $g_{1D}$ for a large 3D scattering length, 
changing completely the physics of the quasi-1D dipolar condensates. 
For polar molecules the dipole strength may be modified by controlling the 
orienting electric field, and the DDI can be scanned from 
zero to large values. The combination of this control and the 
modification of the induced CIR, should allow to scan 
close to the CIR all physics ranging from an homogeneous 
stable quasi-condensate, a bright-solitonic solution, 
a BEC with (eventually unstable) roton minimum, and even a super-Tonks regime.

\acknowledgements
Conversations with H. P. B\"uchler, P. Pedri, D. Petrov, T. Pfau, and G. V. Shlyapnikov, and the support of  
the DFG (SFB-TR21, SFB407, SPP1116) are acknowledged.


\begin{thebibliography}{99}
\bibitem{footnote0} For a repulsive DDI along $x$, the asymptotic
 solution is given by modified Bessel functions $I_1$ and $K_1$, which also
 lead to a logarithmic contribution at large distances.
\bibitem{footnote-PT} We have consider as well a a P\"osch-Teller potential obtaining a similar qualitative behavior.
\end{thebibliography}
\end{document}